\documentclass[showpacs,preprintnumbers,amsmath,amssymb]{revtex4}

\usepackage{graphicx}
\usepackage{dcolumn}
\usepackage{bm}
\usepackage{natbib}
\usepackage{color}

\begin{document}


\title{Statistical Properties of the Inter-occurrence Times in the Two-dimensional Stick-slip Model of Earthquakes}

\author{Tomohiro Hasumi}
 \email{t-hasumi.1981@toki.waseda.jp}
\author{Yoji Aizawa}
\email{aizawa@waseda.jp}
\affiliation{Department of Applied Physics, Advanced School of Science and Engineering, Waseda University, 169-8555, Tokyo, Japan}

\date{\today}

\begin{abstract}
We study earthquake interval time statistics, paying special attention to inter-occurrence times in the two-dimensional (2D) stick-slip (block-slider) model. Inter-occurrence times are the time interval between successive earthquakes on all faults in a region. We select stiffness and friction parameters as tunable parameters because these physical quantities are considered as essential factors in describing fault dynamics. It is found that inter-occurrence time statistics depend on the parameters. Varying stiffness and friction parameters systematically, we optimize these parameters so as to reproduce the inter-occurrence time statistics in natural seismicity. For an optimal case, earthquakes produced by the model obey the Gutenberg-Richter law, which states that the magnitude-frequency distribution exhibits the power law with an exponent approximately unity.
\end{abstract}

\maketitle

\section{Introduction}
Earthquakes are caused by a fracture and frictional slip process. We can understand qualitatively how an earthquake occurs on the basis of the plate tectonic theory proposed by A. Wegener. Statistical properties of earthquakes are well known as empirical laws~\cite{Main:1996}, while the source mechanism of earthquakes is still an open question. For example, Gutenberg and Richter proposed the relation between magnitude ($M$) and frequency ($n$) expressed by
\begin{equation}
\log_{10} n = a - bM,  
\end{equation}
where $a$ and $b$ are positive constants. This relation is called the Gutenberg-Richter (GR) law\cite{Gutenberg:1956}. $b$ is the so-called $b$-value and similar to unity. Strictly speaking, $b$ depends on fault structures and seismicity and ranges from 0.8 to 1.2~\cite{Frohlich:1993}. \par
In general, earthquakes can be categorized into three types: foreshocks, mainshocks, and aftershocks. A mainshock is a large earthquake, whereas a foreshock (aftershock) is an earthquake before (after) the mainshock and which occurred near the mainshock epicenter. Aftershocks obey the Omori law~\cite{Omori:1894}, which stresses that the decay rate of aftershocks follows the power law. Subsequently, a modified version was proposed by Utsu~\cite{Utsu:1961}. Since the Gutenberg-Richter law and the Omori law exhibit the power law, earthquakes are seemed to be self-organized critical phenomena~\cite{Bak:1988, Bak:1989}.\par
\begin{figure*}[t]
\begin{center}
\includegraphics[width=.65\linewidth]{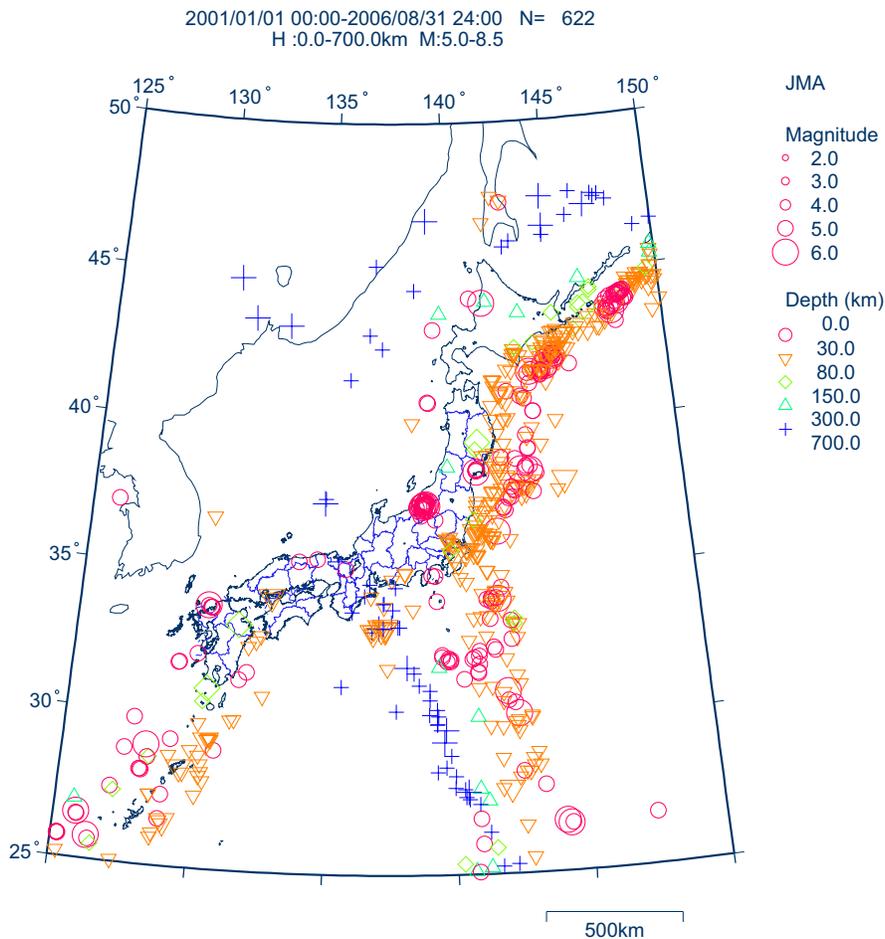}
\caption{Seismicity map around the Japan from 2001/01/01 to 2006/08/31 for $M>5.0$. This map is programed on the basis of the JMA earthquake catalogs.}
\label{map}
\end{center}
\end{figure*}

The time intervals between earthquakes can be classified into two types: recurrence times and inter-occurrence times. Recurrence time is the interval of time between earthquakes on a single fault or segment, whereas inter-occurrence time is the time interval between earthquakes on all faults in a region. For inter-occurrence time statistics, probability distributions have been studied by different authors~\cite{Bak:2002, Corral:2004, Abe:2005} by using earthquake catalogs (see fig.~\ref{map}). Recurrence times are generally used by seismologists to describe the time interval between characteristic earthquakes~\cite{String:1996}. A characteristic earthquake is a large earthquake happening on a single fault and depending on fault length, crust structure, and so on. For recurrence time, several probability distributions have been proposed, such as the log normal distribution, the Weibull distribution, and the double exponential distribution \cite{Matthews:2002}. However, we cannot decide which distribution is the best owing to the lack of data.\par
In this work, we focus on the inter-occurrence time statistics. Statistical properties based on earthquake models have been investigated numerically and compared with seismicity in nature. Then, earthquake models have been modified so as to reproduce fault systems~\cite{Rundle:2000}. Generally, numerical simulations have the advantage of enough earthquakes having occurred to guarantee statistical accuracy. In addition, it is easy to study the statistical properties under various crust conditions by changing control parameters. Optimizing or restricting control parameters so as to adequately reproduce the statistical properties, we suggest the probability distribution function of the recurrence time and offer new insights into earthquake statistics.\par
The stick-slip model proposed by Burridge and Knopoff \cite{Burridge:1967} is often called the block-slider model or the Burridge-Knopoff model. This model describes the relative motion of faults. Although the model is highly simplified, it has been shown that it can extract the statistical properties of earthquakes, such as the GR law, the Omori law, the empirical law of the stress distribution, the constant stress drop, and the inter-occurrence time statistics~\cite{Carlson:1989a, Carlson:1989b, Carlson:1991b, Carlson:1991a, Carlson:1991d, Kumagai:1999, Preston:2000, Mori:2006, Omura:2007, Abaimov:2007, Hasumi:2007}. This model has been modified in order to describe real crust structures \cite{Nakanishi:1992, Hainzl:1999, Xia:2005}. However, the statistics of time intervals in the stick-slip model has not been discussed fully. The purpose of this work is to reveal whether the 2D stick-slip model can be understood as useful model in view of the inter-occurrence time statistics. Thus, we report numerical investigation of earthquake inter-occurrence time statistics produced by the 2D stick-slip model. In this work, stiffness and friction parameters are set as control parameters. Then we restrict or optimize these parameters so as to reproduce the inter-occurrence time statistics in nature. It is concluded that the model reproduces the inter-occurrence time statistics as well as the GR law in a limited parameter regime. Investigating the 2D stick-slip model in a optimal case further, we may propose new findings concerning the statistical properties of earthquakes.

\section{\label{model}Two-dimensional Stick-slip Model}
\begin{figure*}[b]
\begin{center}
\includegraphics[width=.75\linewidth]{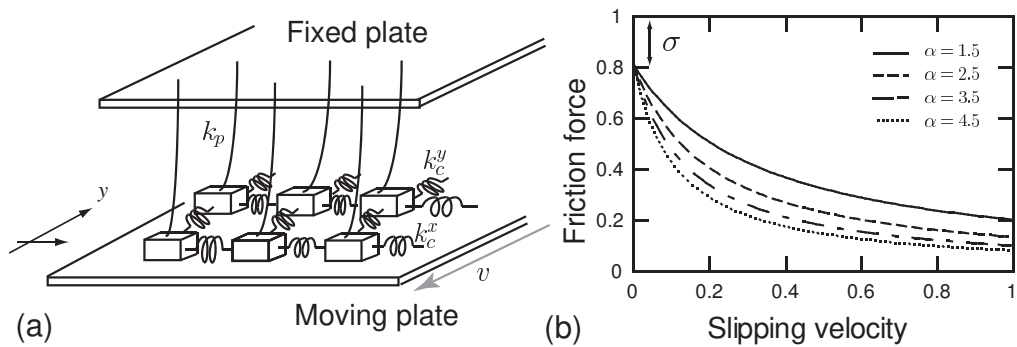}
\caption{(a) 2D stick-slip model. $k_c^x$, $k_c^y$, and $k_p$ are spring constants. The friction force acts on the surface between the block and the bottom plate. (b) $\alpha$-dependence of the non-linear dynamical friction function. $\sigma$ is fixed at 0.01 throughout the simulation.}
\label{BK_dim2}
\end{center}
\end{figure*}
In this work, we numerically investigate statistical properties of the inter-occurrence time, produced by the two-dimensional (2D) stick-slip model. As shown in Fig.~2 (a), the model is composed of blocks on a square lattice, two plates, and two different kinds of springs. The upper plate is fixed, whereas the bottom plate moves at a velocity of $v$. So, this model represents the relative motion of faults. A block corresponds to a segment of a fault, so that we define a one-block-slip event as a minimum earthquake. Shear stresses and compression stress are modeled respectively by coil springs, $k_c^x, k_c^y$ interconnected by a block, and by the leaf springs, $k_p$ between a block and the fixed plate. We set the $-y$ axis as the direction of the loading plate, and the $x$ axis as the perpendicular to the $y$ axis. Assuming that the slip direction of a block is restricted by the $y$-direction only, the block is described by a stick-slip motion. The stick-slip motion can be divided in two parts: one is a stick-state and the other is a slip-state. In the case of the stick-state or loading-state, all blocks and the loading plate move together, whereas a block slips on the bottom plate, the slip state. \par 
The equation of motion in a scaled form at cite $(i,j)$ can be expressed by  
\begin{eqnarray}
\frac{d^2 U_{i,j}}{dt'^2} = l_x^2(U_{i+1,j}+U_{i-1,j}-2U_{i,j}) +  l_y^2(U_{i,j-1}+U_{i,j+1}-2U_{i,j})
 - U_{i,j} -  \phi \left(2\alpha \left(\nu+\frac{dU_{i,j}}{d t'}\right)\right),
\label{eqm_dim2_2}
\end{eqnarray}
where $U, t',$ and $\phi$ correspond respectively to a normalized displacement, time, and a dynamical friction force which is a function of block velocity. Additionally,
\begin{eqnarray}
l_x^2 = \frac{k_c^x}{k_p}, \; \; l_y^2 = \frac{k_c^y}{k_p}, \; \; \nu = \frac{v}{\hat{v}}, \; \; 2\alpha = \frac{\hat{v}}{v_1}, \nonumber
\end{eqnarray}
where $\hat{v}$ is a maximum of the slipping velocity and $v_1$ is the characteristic velocity.
The blocks are subject to the friction acting on the surface between the block and the loading plate. In this study, we adopt \textquotedblleft velocity-weakening" type friction law as a dynamical friction force, $\phi$. This friction law states that a dynamical friction force decreases as the slipping velocity increases. This friction property can be observed in rock-fracture experiments \cite{Scholz:2002} and is formulated mathematically by Carlson {\it et al.}~\cite{Carlson:1991b}, namely,
\begin{eqnarray}
\phi (\dot{U}) = \left\{
\begin{array}{ll}
(-\infty, 1] & \dot{U}=0,\\
{\displaystyle \frac{(1-\sigma)}{\{1+2\alpha[\dot{U}/(1-\sigma)]\}}} & \dot{U}>0.
\end{array}
\right.
\label{friction_function_2}
\end{eqnarray}
It is easy to simulate this friction law so that this formulation has been often used in previous works~\cite{Carlson:1991d, Kumagai:1999, Mori:2006, Hasumi:2007, Nakanishi:1992, Xia:2005}. The friction function can be characterized by two parameters, $\sigma$ and $\alpha$. $\sigma$ is the difference between the maximum friction force $(=1)$ and the dynamical friction force at $v=0~(=\phi(0))$. $\alpha$ is represented the decrement of the friction force, $\phi$.  If $\alpha=0$, $\phi$ is constant, $1-\sigma$. When $\alpha \rightarrow \infty$, $\phi$ decreases quickly to 0. $\nu$ is the normalized plate velocity and is very small parameter. Thus, we set $\nu=0$ when an event happens. This assumption guarantees the condition that no other event occurs during an ongoing event. \par
In this study, we place $50\times 50$ blocks on the $(x,y)$ plane and simulate equation (\ref{eqm_dim2_2}) and (\ref{friction_function_2}) under the free boundary condition by using the 4th-order Runge-Kutta method. Initial configurations of all blocks have small irregularities. $10^5$ order of events after some periods when the initial randomness effect cannot be influenced are used. We study the inter-occurrence time statistics by selecting $l_x, l_y$ and $\alpha$ for tunable or control parameters, while $\nu=0.01$ and $\sigma=0.01$. This work is another version of the previous reports~\cite{Hasumi:2007}.

\section{Results and Discussion}
In this work, the slip of a block is considered as an earthquake. An earthquake occurs when a block slips for the first time during an event. The inter-occurrence time is defined as the time interval between successive events. For example, the $n$th inter-occurrence time can be described by $\tau_n = t'_{n+1}-t'_{n}$, where $t'_{n}$ and $t'_{n+1}$ are the $n$th and the $n+1$th earthquake occurrence time, respectively. In order to compare our results with natural earthquake inter-occurrence time, we introduce scaled inter-occurrence time, $\tau' = \tau/\tilde {\tau}$, where $\tilde {\tau}$ is the normalized scaled time. $\tilde {\tau}$ is set at 1.0 for the model analysis and at 1000 [s] for observation analysis. 

\subsection{Probability density distribution}
\begin{figure}[t]
\includegraphics[width=.75\linewidth]{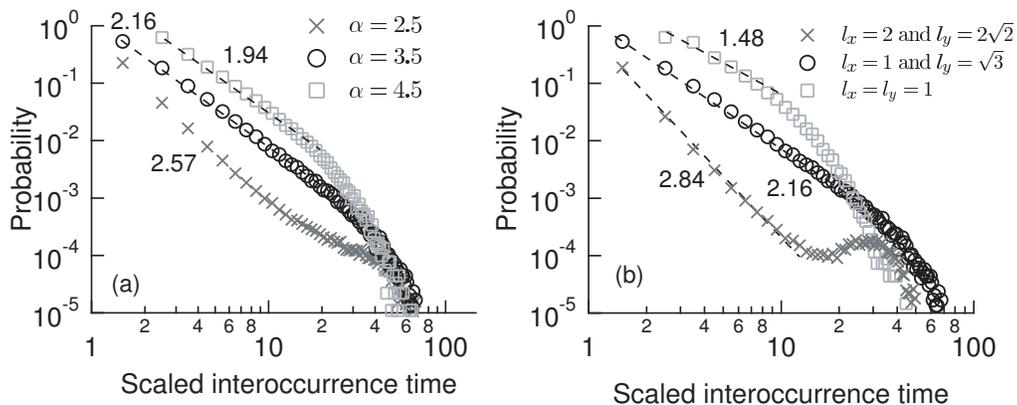}
\caption{The probability distributions of the inter-occurrence time for different control parameters, $l_x,l_y$, and $\alpha$ as a function of the scaled inter-occurrence time. For (a), $\times~(\alpha=2.5)$, $\circ~(\alpha=3.5)$, and $\square~(\alpha=4.5)$, while $l_x=1$ and $l_y=\sqrt{3}$. For (b), $l_x=l_y=1$, $l_x=1$ and $l_y=\sqrt{3}$, and $l_x=2$ and $l_y=2\sqrt{2}$, denoted respectively, $\times$, $\circ$, and $\square$. $b'$ is calculated from the slope of the dash line. All plots except for in the case of $\alpha=3.5$ in (a) and $l_x=1$ and $l_y=\sqrt{3}$ in (b) are shifted vertically for clarity.}
\label{prob}
\end{figure}
Probability density distributions of the inter-occurrence time, $p(\tau')$ are displayed in Fig.~\ref{prob}. For (a), the friction parameter $\alpha$ is changed from 2.5 to 4.5, whereas stiffness parameter is fixed at $l_x=1$ and $l_y=\sqrt{3}$. For (b), $l_x$ and $l_y$ are varied when $\alpha=3.5$. As shown these figures, probability distributions of inter-occurrence time exhibit the power law in short time scale region, $1\lessapprox \tau \lessapprox 7$. We calculate the power law exponent, denoted here $\beta$ from the slope of the distribution; for example, 2.40 $(\alpha=2.5)$, 1.96 $(\alpha=3.5)$, 1.78 $(\alpha=4.5)$ for (a). In the case of $\alpha=3.5$ the distribution shows the power law, $1\lessapprox \tau \lessapprox 20$, so that the system exhibits a critical state approximately (type A). For $\alpha=2.5$, the probability of a long inter-occurrence time region, $\tau \gtrapprox 10$, is enhanced more than expected by the power law decay, hereafter referred to as type B. On the other hand, as when $\alpha=4.5$, the probability is less than predicted by the power law~(type C). As for (b), $\beta$ increases when $l_x$ and $l_y$ are enhanced, such as $\beta \simeq 1.94$, 1.96, and 2.71 for  $(l_x = l_y = 1)$,  $(l_x=1$ and $l_y = \sqrt{3})$, and  $(l_x=2$ and $l_y = 2\sqrt{2})$DThe forms and trends of the distributions can be categorized into three types, type A $(l_x=1$ and $l_y = \sqrt{3})$, type B $(l_x=2$ and $l_y = 2\sqrt{2})$, and type C $(l_x = l_y = 1)$. It is found that the form of the distribution in a long time region and the power law exponent $\beta$ depend on the dynamical parameters, $l_x,l_y$, and $\alpha$. Our findings are different from those of previous works studying of the 1D stick-slip model~\cite{Preston:2000,Mori:2006,Omura:2007,Abaimov:2007}. 

\subsection{Survivor function}
\begin{figure*}[t]
\includegraphics[width=.75\linewidth]{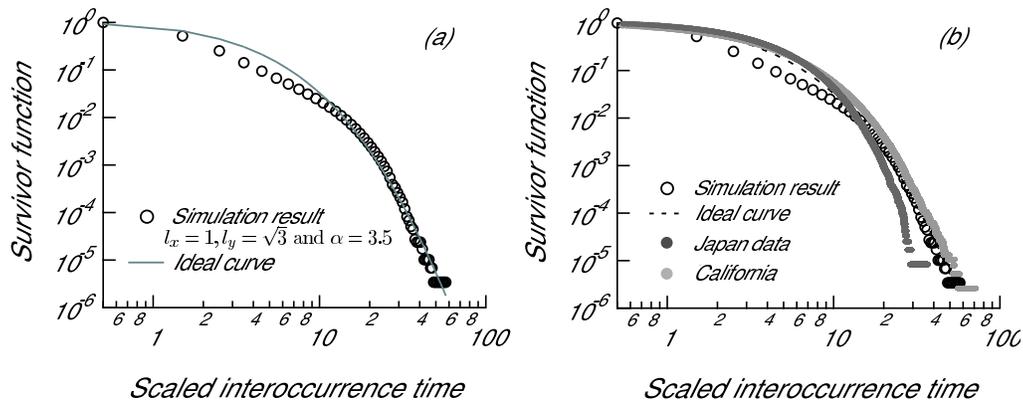}
\caption{The survivor functions of inter-occurrence time statistics. For (a), the simulation is performed $l_x=1, l_y=\sqrt{3}$, and $\alpha=3.5$. For (b) the survivor distributions for the optimal case of the model, Japan earthquakes, and California earthquakes are shown. The fitting parameters are calculated as $q=1.06$ and $\tau_0=2.55$ for the model, $q=1.05$ and $\tau_0 = 3.16$ for Japan, and $q=1.13$ and $\tau_0=3.44$ for California.}
\label{survivor}
\end{figure*}
Up to now, we have been discussing statistical properties of the survivor function, $D(\tau) = prob~(t>\tau) = 1-F(\tau)$, where $F(\tau)$ is the cumulative distribution. Abe and Suzuki have analyzed the Japan and Southern California earthquake catalogs and found that the survivor function of the inter-occurrence time can be described by the power law as~\cite{Abe:2005}
\begin{eqnarray}
D(\tau') = \frac{1}{(1+\epsilon \tau)^\gamma},
\end{eqnarray}
where $\gamma$ and $\bar{\tau}$ are parameters. This distribution function can be written as
\begin{eqnarray}
D(\tau') = e_q(-\tau/\tau_0)= [\left(1+(1-q)(-\tau/\tau_0)\right)^{\frac{1}{1-q}}]_{+}, \; ([a]_{+} \equiv \textrm{max} [0,a]),
\label{ZM}
\end{eqnarray}
where $q$ and $\tau_0$ are positive constants and are related to $\gamma$ and $\epsilon$: $\gamma= 1/(q-1)$ and $\epsilon = (q-1)/\tau_0$. $e_q(x)$ is the so-called $q$-exponential distribution derived from the non-additive statistical mechanics proposed by Tsallis~\cite{Tsallis:1988}. In this work, we select the power law distribution defined by Eq.~(\ref{ZM}) for the ideal survivor distribution function, and then optimize the control parameters, $l_x, l_y$ and $\alpha$.\par

We present the survivor function of the inter-occurrence time in the case of $l_x= 1, l_y = \sqrt{3}$, and $\alpha=3.5$ in Fig.~\ref{survivor} (a). Plots and the dashed line correspond to the numerical data and the ideal curve, respectively. Varying the control parameters, $l_x, l_y$, and $\alpha$ systematically, we can find the optimal case of the control parameters: $l_x= 1, l_y = \sqrt{3}$, and $\alpha=3.5$. For an optimal case, the fitting parameters of the survivor functions are estimated to be $q=1.06, \tau_0=2.55$ and the correlation function $\rho$ yields 0.986. Figure~\ref{survivor} (b) shows the inter-occurrence time statistics obtained from the model (optimal case), Japan earthquakes, and California earthquakes. Therefore, it is concluded that the inter-occurrence time can be reproduced semi-qualitatively for the optimal case. \par

\begin{figure*}[b]
\includegraphics[width=.75\linewidth]{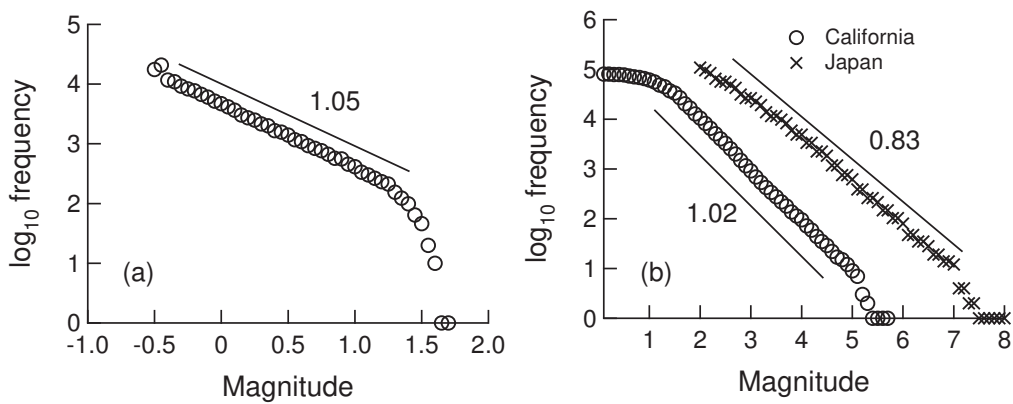}
\caption{The magnitude distribution obtained from the model in (a) and from real earthquakes in (b). The $b$-value of the GR law is calculated from the slope of the distribution: 1.05 (model), 0.83 (Japan), and 1.02 (California).}
\label{GR law}
\end{figure*}
Finally, magnitude distributions derived from the model in the case for $l_x= 1, l_y = \sqrt{3}$, and $\alpha=3.5$ and from Japan and California earthquake catalogs are shown in Fig.~\ref{GR law} (a) and (b), respectively. Note that we use the JMA catalogs: \textquotedblleft http://kea.eri.u-tokyo.ac.jp/tseis/jma1/" for Japan and the NCEDC catalogs: \textquotedblleft http://www.ncedc.org/ncedc/catalog-search.html" for California for the period 2001/01/01-2006/08/31. It should be noted that the JMA catalog lacks earthquake data whose magnitude is less than 2.0. Comparing Fig.~\ref{GR law} (a) with (b), we demonstrate that the model can reproduce the GR law, which we explained previously: the power law distribution with exponent, $b$-value 1.0. We found that in the case of optimal parameters, the inter-occurrence time statistics and the GR law can be extracted from the 2D stick-slip model~\cite{Hasumi:2007}.

\section{Conclusion}
In this study we numerically investigated the statistics of earthquake interval times, the inter-occurrence time based on the 2D spring-block model. Inter-occurrence times are interval times between earthquakes on all faults in a region. It is found that inter-occurrence time statistics depend on the control parameters, $l_x, l_y$, and $\alpha$ charactering the model. $l_x$ and $l_y$ are stiffness parameters, whereas $\alpha$ expresses the decrement of dynamical friction. The probability density distributions of the inter-occurrence time show the power law in the short-time region. For the long time region, the distributions could be classified into three types: power law behavior (type A), broad peak structures (type B), and exponential cutoffs (type C). Then, we restricted the control parameters so as to reproduce the inter-occurrence time statistics in nature; the survival function of the inter-occurrence time revealed the $q$-exponential distribution with $q>1$. The optimal parameters are estimate to be $l_x= 1, l_y = \sqrt{3}$, and $\alpha=3.5$. \par
In the case of the optimal parameters, the magnitude distribution shows the power law with exponent 1.0. This power law distribution is similar to the GR law, and the exponent corresponds to the $b$-value, which is characterized by the GR law. Hence, we demonstrate that in the optimal case, $l_x= 1, l_y = \sqrt{3}$, and $\alpha=3.5$, the model can reproduce the GR law and the inter-occurrence time statistics both simultaneously and spontaneously.\par
We acknowledge that the stick-slip (block-slider) model is highly simplified so that many effects playing important roles in fault dynamics have been neglected. However, it is shown that this model is useful for the extraction of the statistical properties of earthquakes because the inter-occurrence time statistics and the GR law can be extracted. This work is a first step toward studying the origin of the statistics of time intervals. We hope to extend our work by focusing on recurrence time statistics and comparing them with natural recurrence time statistics. 

\section*{Acknowledgments}
T.H. would like to thank Prof. S. Abe, Prof. N. Suzuki,  Prof. H. Kawamura, Dr. M. Kamogawa, Dr. T. Sato, Dr. T. Hatano, Dr. Y. Kawada, and Dr. T. Mori for useful comments, fruitful discussions, and manuscript improvements. This work was partly supported by the Japan Society for the Promotion of Science (JSPS), the Earthquake Research Institute cooperative research program at University of Tokyo, and a grant to the 21st Century COE Program \textquotedblleft Holistic Research and Education Center for Physics of Self-organization Systems" at Waseda University from the Ministry of Education, Culture, Sports, Science, and Technology (MEXT), Japan.


\end{document}